\documentstyle[epsf,rotate,twocolumn,aps,pre]{revtex}

\begin{document}

\twocolumn[\hsize\textwidth\columnwidth\hsize\csname
@twocolumnfalse\endcsname

\title{
Fluids of hard ellipsoids: Phase diagram including a nematic
instability from Percus--Yevick theory}

\author{M. Letz, A.~Latz\\
Johannes-Gutenberg Universit\"at, 55099 Mainz, Germany} 

\date{\today}

\maketitle

\begin{abstract}
An important aspect of molecular fluids is the relation between
orientation and translation parts of the two--particle
correlations. Especially the detailed knowledge of the influence of
orientation correlations is needed to explain and calculate in detail
the occurrence of a nematic phase.\\
The simplest model system which shows both orientation and translation
correlations is a system of hard ellipsoids. We investigate an
isotropic fluid formed of hard ellipsoids with Percus--Yevick theory.\\
Solving the Percus--Yevick equations self-consistently in the high
density regime gives a clear criterion for a nematic instability. We
calculate in detail the equilibrium phase diagram for a fluid of hard
ellipsoids of revolution. Our results compare well with Monte Carlo
Simulations and density functional theory. 
\end{abstract}
\vskip 0.2cm
\hspace*{2cm} pacs: {61.25.Em, 61.30.Cz, 61.20.Gy}
\vskip 0.5cm
]
\narrowtext

\section{Introduction}
\label{sec:intro}
Isotropic simple liquids formed of atomic
systems with rotational symmetry are well understood. 
If the two particle correlation is given by a hard sphere interaction an
integral equation like the Percus--Yevick (PY) \cite{hansen76} 
closure relation can be
solved analytically. For the liquid phase the PY
equation gives good results even in the dense liquid regime 
(up to a packing fraction $\phi < 0.49$)
above which
the equilibrium state is crystalline which PY fails to obtain.
The packing fraction $\phi$ is
defined as the relation between the number density $\rho$ and the
volume of the particles $\phi = \frac{\pi}{6} \rho \sigma^3$ with
$\sigma$ usually set to one. 
For hard spheres $\phi \approx 0.64$ corresponds to
random closed packing and $\phi = \sqrt{2} \pi /6 \approx 0.74$ to
packing in a fcc lattice.\\[0.2cm]  

Molecular systems have usually complicated potentials which are
modeled e.g. by Lennard-Jones potentials of each atom in a
molecule. A very basic feature of molecular systems is 
the existence of orientation degrees of freedom 
which interplay in non--trivial ways with translational degrees of
freedom.
The simplest model
system which allows to study this interplay is a system of 
rotational symmetric hard
ellipsoids. The equilibrium phase diagram is now -- compared to hard spheres --
enriched by an additional variable, the aspect ratio $X_0$ of the
ellipsoids, which is defined as the ratio between the major axis, a,
and the minor axis, b, $X_0 = \frac{a}{b}$.
As a function of the aspect ratio or density a fluid of hard
ellipsoids can now also undergo an isotropic to nematic (I-N) transition.
This is expected from the Onsager solution of hard spherocylinders
\cite{onsager49}, 
has been found by computer simulations \cite{frenkel84} and also
by density functional theory \cite{groh97}.\\[0.2cm]

PY theory deals with the isotropic phase. It is not able to describe a
phase transition. However it is known from the hyper-netted--chain
(HNC) closure relation for hard ellipsoids \cite{perera87} and for
dipolar hard spheres \cite{klapp96} that it is in principle possible
to identify a precursor phenomenon of a phase transition in an
integral equation.   
Such a precursor
phenomenon is not known for the PY closure relation (see also
\cite{chamoux96}).  
  
In this paper however we show that the PY closure relation leads 
not only to a
clear indication of a nematic instability but enables us also to
calculate the equilibrium phase diagram of hard ellipsoids.

\section{Integral equations}
\label{sec:inteq}

A fundamental relation which all closure relations for integral
equations are based upon is the Ornstein--Zernike (OZ) equation  \cite{gray84} 
\begin{eqnarray}
\lefteqn{
h({\bf r}_1,\Omega_1,{\bf r}_2,\Omega_2) = 
c({\bf r}_1,\Omega_1,{\bf r}_2,\Omega_2)}\nonumber \\  &\;\;\;\;\;+& \rho 
\Big ( c({\bf r}_1,\Omega_1,{\bf r}_3,\Omega_3) 
h({\bf r}_3,\Omega_3,{\bf r}_2,\Omega_2) \Big )_{{\bf
r}_3,\Omega_3}\;\;\; .
\end{eqnarray}
It provides the relation between the total correlation function
$h({\bf r}_1,\Omega_1,{\bf r}_2,\Omega_2)$ and the direct correlation function
$c({\bf r}_1,\Omega_1,{\bf r}_2,\Omega_2)$. 
The product is given by
\begin{equation}
\Big ( ...  \Big )_{{\bf r}_3,\Omega_3} = 
\frac{1}{16 \pi^2} \int d\Omega_3 \int d{\bf r}_3 ... \;\;\;\;\; ,
\end{equation}
${\bf r}_i$ is the position of the center of mass of the ellipsoid $i$
and $\Omega_i$ is the orientation of this ellipsoid represented by the
Euler angles $\Phi_i, \theta_i$
(the third Euler angle $\chi$ is not needed due to the symmetry of the
ellipsoids).
Due to translational invariance the
functions depend on ${\bf r}_{ij} = {\bf r}_i - {\bf r}_j$ only and
due to the isotropy (in the liquid) the correlation functions only
depend on $r = |{\bf r}_{ij}|$
once a specific coordinate system has been chosen.
In the following subsections we transform
the OZ equations and the PY closure relation (eq. (\ref{eq:py1})) in 
such a way that we can use it for a numerical treatment. With most of
the definitions we follow the book of Gray and Gubbins \cite{gray84}. 

\subsection{Spherical harmonics}
An obvious orthogonal basis set to expand the angular dependence of the
correlation functions are spherical harmonics. The transformed
correlation function $F \in \{ c,h, ...\}$ is given by:
\begin{eqnarray}
\label{eq:sph}
\lefteqn{
F(l_1,l_2,m;r) } \nonumber \\ &&= \;\; i^{(l_1-l_2)}
\int d \Omega_1 \int d \Omega_2 F(\Omega_1, \Omega_2,
r) Y^m_{l_1}(\Omega_1) Y^{m\;*}_{l_2}(\Omega_2)
\end{eqnarray}
Note that our transformation differs by a factor of $(-1)^m$ and by a
factor of $i^{(l_1-l_2)}$ from the definition used in the book of Gray
et al. \cite{gray84}. The latter of these two factors gives us (in
$q$-frame) only real elements of the correlation functions.
For eq. (\ref{eq:sph}) the r-frame was used.
This means the z--axis of the
coordinate system 
was chosen along the axis connecting the two particles which are
correlated \cite{gray84}. Therefore we only need to deal with one index $m$.
In $q$-space we use, after Fourier transformation the
laboratory fixed frame, 
the $q$-frame \cite{cummings83,gray84} where now all $q$-dependent
correlation functions are diagonal in m.
Within the complete set of spherical harmonics the OZ--equation
can be rewritten:
\begin{eqnarray}
\label{eq:ozsh}
\lefteqn{
h(l_1,l_2,m;r) = c(l_1,l_2,m;r) }
\nonumber \\ & \;\;\;\; + &
\frac{\rho}{4 \pi} \sum_l \int dr_1 \;\;\;
 c(l_1,l,m;r_1) h(l,l_2,m;r-r_1)
\end{eqnarray}
This equation relates the total correlation function $h(l_1,l_2,m;r)$
with the direct correlation function $c(l_1,l_2,m;r)$. The total
correlation function has two contributions, a direct one which results
from direct correlations and is just $c(l_1,l_2,m;r)$
plus an indirect contribution which averages
over possible interactions mediated by another particle in an indirect
way.

\subsection{Transformation into q-space}
Due to the expansion 
in spherical harmonics a Fourier transform cannot be performed as
usual. First one has to find a representation of $F$ which is
invariant with respect to rotation.
\begin{eqnarray}
\label{eq:invh}
\lefteqn{
F(l_1,l_2,l;r) = } \nonumber \\ & \;\;\; &
\sum_m \sqrt{\frac{4 \pi}{(2l+1)}} \;
F(l_1,l_2,m;r) \; C(l_1,l_2,l;m,-m,0)
\end{eqnarray}
where $C(l_1,l_2,l;m_1,m_2,m)$ are the the Clebsch--Gordon coefficients. 
The next step is the 
Hankel transformation which uses the
Rayleigh expansion to transform $F$
from r--space to q--space. This involves spherical Bessel functions
$j_l(qr)$ due to the expansion of
$e^{iqr}$ within the basis of spherical harmonics.
\begin{equation}
\label{eq:reigh}
F(l_1,l_2,l;q) = 4\pi (-i)^l \int_0^{\infty} j_l(qr) \; F(l_1,l_2,l;r) 
\end{equation}
In the final step one goes from the rotational invariant 
representation to the q-frame 
and one
gets a representation of $F$.
\begin{eqnarray}
\label{eq:invz}
\lefteqn{
F(l_1,l_2,m;q) = } \nonumber \\ & \;\;\; &
\sum_l \sqrt{\frac{(2l+1)}{4 \pi}} \;
F(l_1,l_2,l;q) \; C(l_1,l_2,l;m,-m,0)
\end{eqnarray}
Therefore the equations (\ref{eq:invh}) - (\ref{eq:invz}) transform a
two particle correlation function given in real space and $r$-frame into
a function in $q$-space and $q$-frame.

\subsection{The Ornstein Zernike equations in q-space}
Applying eqs. (\ref{eq:invh}) - (\ref{eq:invz}) to the OZ
equation one can rewrite eq. (\ref{eq:ozsh}):
\begin{eqnarray}
\lefteqn{
h(l_1,l_2,m;q) = c(l_1,l_2,m;q)}
\nonumber \\ & \;\;\;\;  &
 + \frac{\rho}{4 \pi} \sum_l 
c(l_1,l,m;q) h(l,l_2,m;q)
\end{eqnarray}
This can be written as a matrix equation for each $m$ and $q$ value:
\begin{equation}
\underline{\underline{h}}(m;q) = \underline{\underline{c}}(m;q) 
+ \frac{\rho}{4 \pi}
\underline{\underline{c}}(m;q) \underline{\underline{h}}(m;q)
\end{equation}
$\underline{\underline{c}}, \underline{\underline{h}}$ are symmetric
matrices with indices $l_1, l_2$. 

For the input into our numerical calculation we define an auxiliary
correlation function $y$ in the usual
way \cite{gray84}: 
\begin{equation}
\underline{\underline{y}}(m;q) = \underline{\underline{h}}(m;q) - 
\underline{\underline{c}}(m;q)
\end{equation}
Using this auxiliary function the OZ equation rewrites:
\begin{equation}
\label{eq:oz}
\left ( 1-\frac{\rho}{4 \pi} \underline{\underline{c}}(m;q) \right ) 
\underline{\underline{y}}(m;q) = \frac{\rho}{4 \pi} \left [
\underline{\underline{c}}(m;q) 
\right ] ^2
\end{equation}
This is a linear system of equations which determines $y(l_1,l_2,m;q)$
if $c(l_1,l_2,m;q)$ is known.

\section{Percus-Yevick closure relation}

In a formal way one can define a product between two correlation
functions $c = a \; * \; b$ \cite{gray84a}. In
r-frame this product reads:
\begin{eqnarray}
\label{eq:py}
\lefteqn{
c(l_1,l_2,m;r) = \frac{1}{4 \pi} \sum_{ l_1' l_2' \atop  l_1'' l_2''}
\sqrt{
\frac{(2l_1'+1)(2l_2'+1)(2l_1''+1)(2l_2''+1)}{(2l_1+1)(2l_2+1)}
}} \nonumber \\ &\;\;\;&
C(l_1',l_1'',l_1;0,0,0) \; C(l_2',l_2'',l_2;0,0,0)
\nonumber \\ &\;\;\;&
\sum_{m',m''}
C(l_1',l_1'',l_1;m',m'',m) \; C(l_2',l_2'',l_2;-m',-m'',-m)
\nonumber \\ &\;\;\;&
a(l_1',l_2',m';r) \; b(l_1'',l_2'',m'';r)
\end{eqnarray}
The Clebsch--Grodan coefficients enter into the equation due to
spatial rotations which have to be performed.

The PY closure relation can now be expressed as 
\begin{equation}
\label{eq:py0}
c = b \; * \; g
\end{equation}
were $g$ is the pair correlation function and via $b=1-e^{\beta u}$
the pair potential $u$ enters into the equation. For the purpose of
solving the PY equation numerically it is better to rewrite
eq. (\ref{eq:py0}) as a function of the auxiliary function $y$.
\begin{equation}
\label{eq:py1}
c = f \; * \; y
\end{equation}
were $f$ is the Mayer function 
\begin{equation}
f(\Omega_1,\Omega_2,r) = e^{-\beta \; u(\Omega_1,\Omega_2,r)} - 1
\end{equation}
The matrix elements of $f$ in the basis set of spherical harmonics
have to be computed using eq. (\ref{eq:sph}).\\[0.2cm]
This equation (\ref{eq:py1}) determines the direct correlation function
$c$ if the 
auxiliary function $y$ and the Mayer function $f$ are known.

\subsection{The pair potential}
In order to determine the matrix elements of the Mayer function 
\begin{equation}
\label{eq:flm}
f(\Omega_1,\Omega_2,r) = \left \{
\begin{array}{ccc}
0 & for & D(\Omega_1,\Omega_2,r) < r \\
-1 & for & D(\Omega_1,\Omega_2,r) \ge r 
\end{array} \right .
\end{equation}
we use
the well known approximation of Berne and Pechukas \cite{berne72},
were $D$ depends on the relative orientation of the two ellipsoids:
\begin{eqnarray}
D(\Omega_1,\Omega_2,r) &=& \left [
1 - \frac{1}{2} \chi \left ( 
\frac{(\cos\theta_1 + \cos\theta_2)^2}{1 + \chi \; ({\bf e}_1  {\bf
e}_2)} \right . \right . \nonumber \\  &\;\;\;\;\; & \;\;\;\;\;\;\;\; 
+ \left . \left .
\frac{(\cos\theta_1 - \cos\theta_2)^2}{1 - \chi \; ({\bf e}_1  {\bf
e}_2)}
\right ) \right ]^{-1/2}
\end{eqnarray}
${\bf e}_i$ are unit vectors along the symmetry axis of an
ellipsoid on position $i$. This approximation models the interaction
between two ellipsoids by the overlap of Gaussians.
The value of $\chi$ is related to the
aspect ratio of the ellipsoids
\begin{equation}
\label{eq:chi}
\chi = \frac{X_0^2-1}{X_0^2 + 1}
\end{equation}
Note that $D(\Omega_1,\Omega_2,r)$ is {\bf not} invariant under X$_0
\rightarrow $X$_0^{-1}$ which implies $\chi \rightarrow -\chi$.

\subsection{Symmetries of the solution}

Due to the symmetries of the ellipsoid there are certain simplifications in
the calculation which can be applied.
\begin{itemize}
\item[(i)]
Due to the head--tail symmetry of the ellipsoids all matrix elements
of a correlation function $F(l_1,l_2,m,u)$ $u \in \{r,q \}$ with 
$l_i$ odd are zero. 
\item[(ii)]
All elements of $F$ are real both in r-space and q-space.
Using the definition of eq. (\ref{eq:sph}) this is even valid for all
linear molecules.
\item[(iii)]
Therefore there is an additional symmetry $F(l_1,l_2,m,u) =
F(l_1,l_2,-m,u)$
\item[(iv)]
Also the $l$ occurring in the rotational invariants
(eq. (\ref{eq:invh})) can only have even values following from
$l_1+l_2+l = $even which results from inversion symmetry.
\end{itemize}
For a proof of (ii) and (iii) the reader may consult
\cite{schilling97} and for (i) and (iv) one might look up \cite{gray84}.
There is one further feature we want to point out: For
small argument ($r$ or $k$), all non diagonal elements
($l_1 \neq l_2$) have to vanish, at least in the isotropic phase:
\begin{equation}
\label{eq:ndze}
\lim_{u \rightarrow 0} F(l_1,l_2,m,u) = 0 \;\;\; , \;\;\; \mbox{if} \;\;
l_1 \neq l_2
\end{equation}
This follows from the transformation of $F$ under rotations $R$:
\begin{eqnarray}
\lefteqn{
\lim_{u \rightarrow 0} F(l_1,l_2,m,Ru) \nonumber } \\ && = 
\lim_{u \rightarrow 0} \sum_{m_1,m_2} D_{m_1,m}^{l_1 \; *}(R)
D_{m_2,m}^{l_2}(R) F(l_1,l_2,m_1,m_2,u)
\end{eqnarray}
This equation has to be valid for all $R$ which results in
$\delta_{l_1,l_2} \delta_{m_1,m_2} $. This can be seen by integrating
both sites of the above equation and by making use of the unitarity if
the rotation matrices:
\begin{eqnarray}
\lefteqn{
\int dR
\lim_{u \rightarrow 0} F(l_1,l_2,m,Ru) \nonumber } \\ && = 
8 \pi^2 F(l_1,l_2,m,0) \nonumber  \\ && = 
\sum_{m_1,m_2} 
\int dR D_{m_1,m}^{l_1 \; *}(R)
D_{m_2,m}^{l_2}(R) F(l_1,l_2,m_1,m_2,0)
\nonumber  \\ && =
\sum_{m_1,m_2} \frac{8 \pi^2}{2l_1+1} \delta_{l_1,l_2}
\delta_{m_1,m_2} 
F(l_1,l_2,m_1,m_2,0)
\end{eqnarray}
Therefore $F$ has to vanish for small $u$ for all non-diagonal elements
($l_1 \neq l_2$).
Further the value of the diagonal
elements of $F(l_1,l_1,m,u \rightarrow 0)$ 
with $(2l_1+1)$ different $m$
can (for short ranged potentials) not depend on $m$. This symmetry can
clearly be seen from 
fig. \ref{fig:cmp}c and from fig. \ref{fig:nem}c.

\subsection{Calculation procedure}
In order to obtain a numerical solution of the equations above the
following steps have to be performed:
\begin{itemize}
\item[1.)]
The matrix elements $f(l_1,l_2,m,r)$ of the Mayer function have to be
computed using eq. (\ref{eq:sph}) and eqs. (\ref{eq:flm}) -
(\ref{eq:chi}). For our calculation we used 100 points in the range
Min$(1,X_0) < r < $ Max$(1,X_0)$ where $r$ is given in units of the
major axis $a$ of the ellipsoids.   
\item[2.)]
An initial guess for $c(l_1,l_2,m,r)$ has to be made and a grid
for $r$ of $N_{max} = 400$ points in the range $0 < r < 10$ was chosen.
\item[3.)]
The iteration begins by 
using eqs. (\ref{eq:invh}) - (\ref{eq:invz}) to go from
$c(l_1,l_2,m,r)$ in r--space and r--frame to $c(l_1,l_2,m,q)$
in q--space and q--frame using the Rayleigh transformation for the
direct correlation function $c(l_1,l_2,m,r)$. It turned out to be
crucial to use an analytic expansion of the spherical Bessel functions
$j_l(x)$ for small argument $x$.
\item[4.)]
The auxiliary function 
$y(l_1,l_2,m,q)$ has to be calculated using the  OZ
eqs. in the form of eq. (\ref{eq:oz}). As a q-space grid we used 400
points in the range $0 < q < 50$ were q is measured in units of $\left
[ \frac{2 \pi}{a} \right ]$.
\item[5.)]
Using eqs. (\ref{eq:invz}) - (\ref{eq:invh}) we get the function 
$y(l_1,l_2,m,r)$ in r--space and r--frame.
\item[6.)]
With the help of the PY eq. (\ref{eq:py}) one can obtain the next
iteration for $c(l_1,l_2,m,r)$.
\item[7.)]
The steps 3.) to 6.) have to be iterated until a fix point of the
equations has been reached with a given accuracy. In this way a
self-consistent solution can be found.
\end{itemize}
As a test for self-consistency we choose the mean square deviation of
$c$ between two steps of iteration
\begin{eqnarray}
\lefteqn{
\epsilon = \frac{1}{(l_{max}-1) (m_{max}+1) N_{max}}\nonumber } \\
&&
\sqrt{
\sum_{l_1,l_2,m,n} \left ( c^{(p+1)}(l_1,l_2,m,r_n) -
c^{(p)}(l_1,l_2,m,r_n) \right )^2 }
\end{eqnarray}
where the summation indices were in the regions
$l_1,l_2 \in \{ 0,2, ... ,l_{max} \}$, $m \in \{ 0,1,2,3, ...,
m_{max} \} $ and at the end $\epsilon$ was typically chosen to be
smaller then $2 \cdot 10^{-5}$ as a condition for convergence.  

In this way one can obtain a stable self-consistent solution for the
correlation functions. This has already been done for a fluid of
ellipsoids 
in refs.
\cite{perera87,pospisil93,ram91,ram94} 
and also in ref. \cite{franosch97b} for a single ellipsoid in a fluid
of hard spheres.
In this work we have extended the
calculation to a much higher density regime than it has been done in
previous works. In order to reach the high densities we 
were forced to restrict 
the maximum numbers for $N_{max}$ was set to $400$ and $l_{max}$
and $m_{max}$ to 4. The value of $l$ in the rotational invariants 
was restricted to
$l \in \{ 0,2,4, ... , 2l_{max} \}$. 

\section{Results from PY}
The virial expansion of hard ellipsoids of revolution is symmetric
with respect to $X_0 \longrightarrow 1/X_0$ up to the second order in
density. 
This approximate
symmetry is violated for 
higher densities.
\begin{figure}
\unitlength1cm
\epsfxsize=8cm
\begin{picture}(7,18)
\put(-1.2,12){\rotate[r]{\epsffile{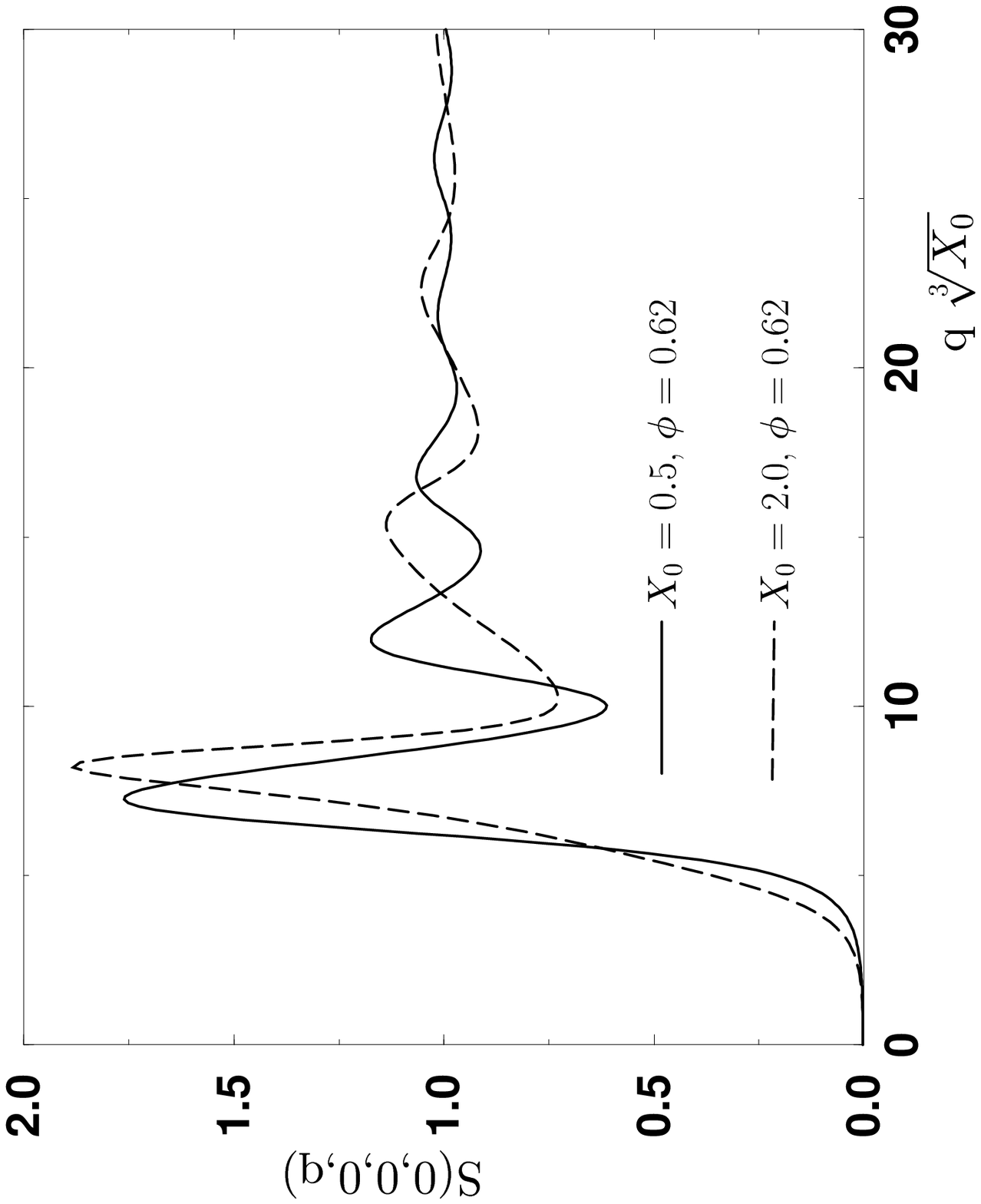}}}
\epsfxsize=8cm
\put(-1.2,6){\rotate[r]{\epsffile{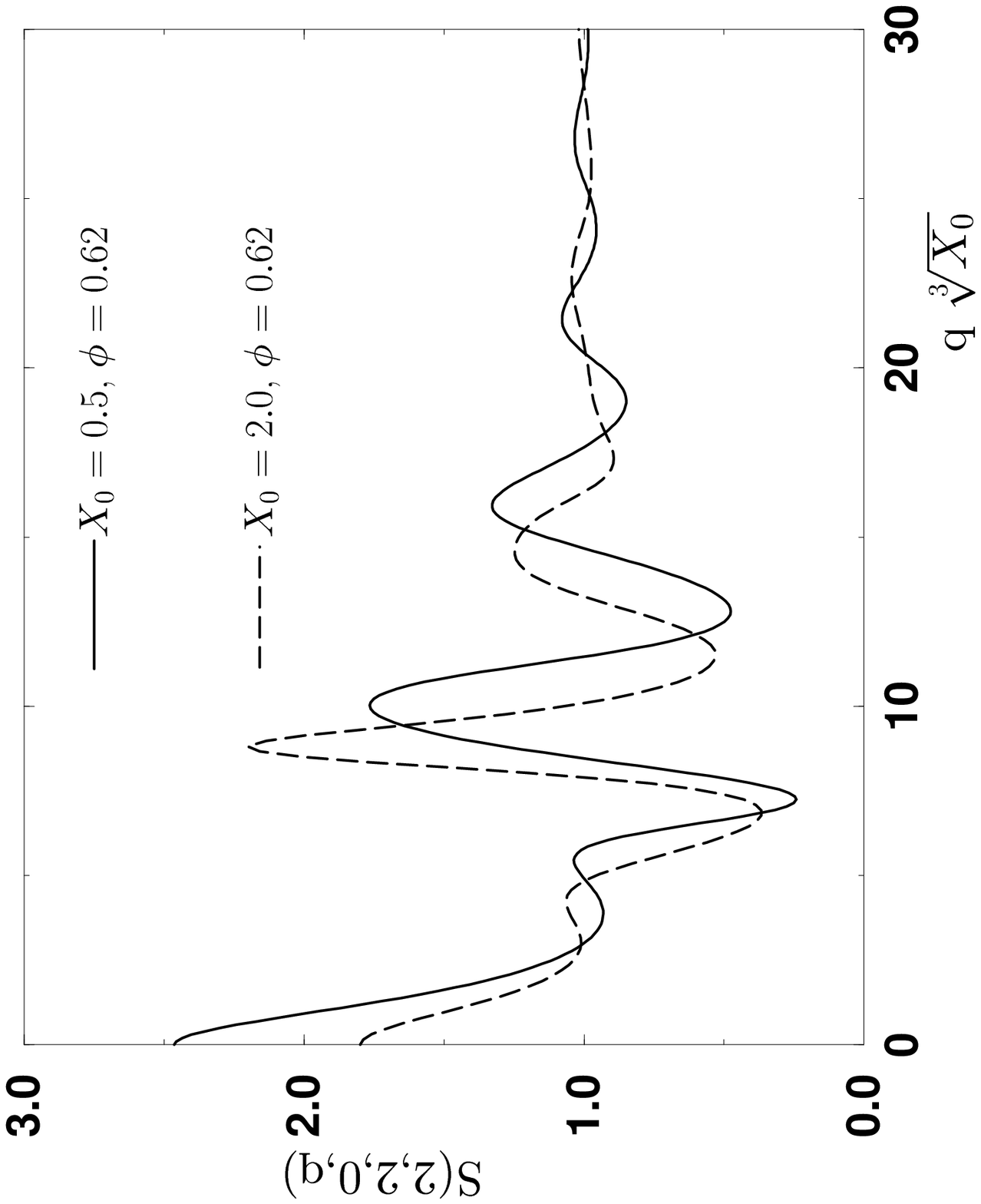}}}
\epsfxsize=8cm
\put(-1.2,0){\rotate[r]{\epsffile{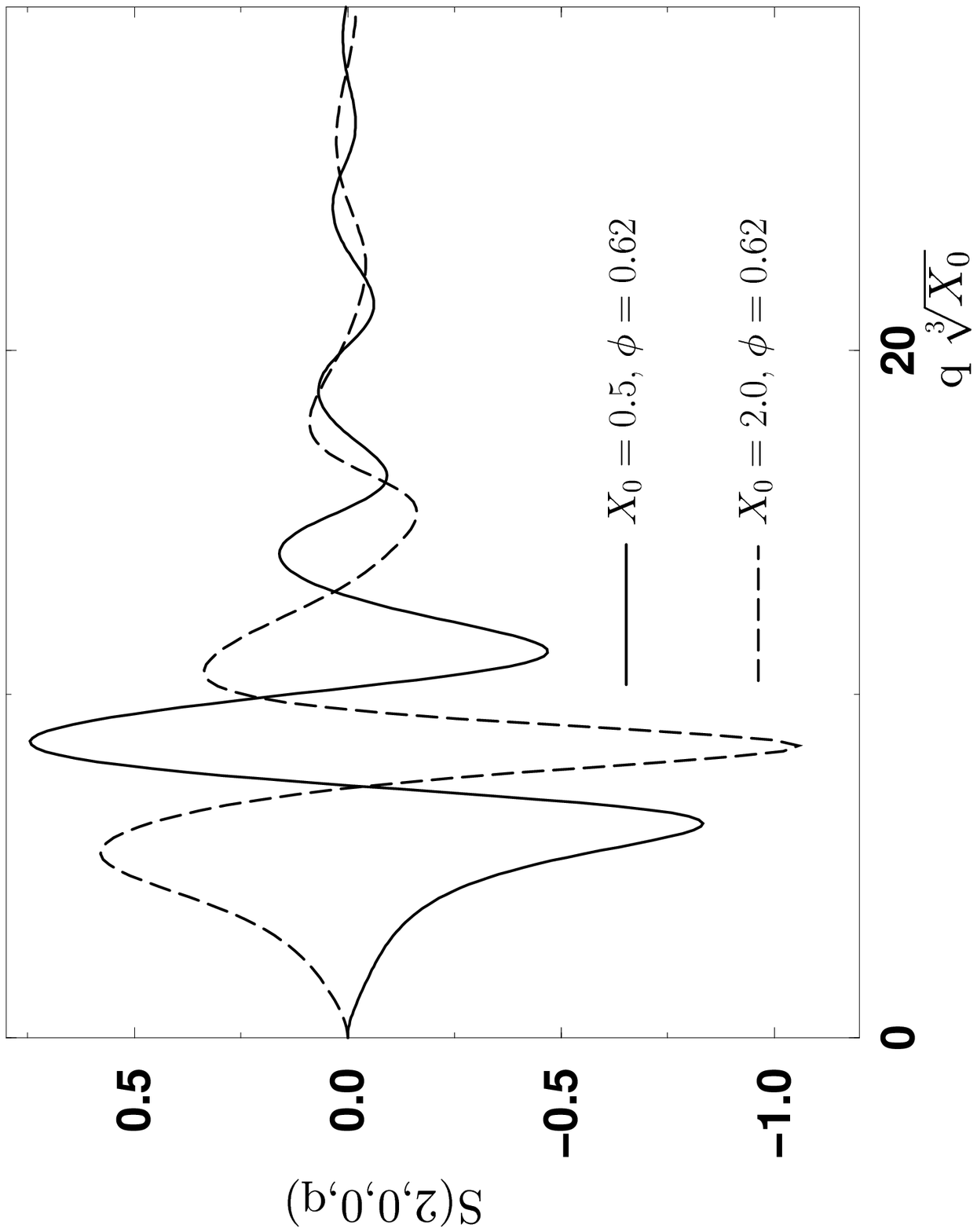}}}
\end{picture}
\caption{{
In order to demonstrate the breakdown of the approximate symmetry
between prolate and oblate ellipsoids ($X_0 \longrightarrow 1/X_0$) we
plotted for $X_0 = 2.0$ and $X_0 = 0.5$ elements of $S(l_1,l_2,m,q)$,
(a) $l_1 = l_2 = m = 0$, (b) $l_1 = l_2 =2,  m = 0$, (c) $l_1 = 2, l_2
= m = 0$
The q-axis was scaled by a factor of 
$ X_0^{1/3} $. 
}}
\label{fig:cmp}
\end{figure}
This is shown in fig. \ref{fig:cmp} where we have plotted 
three matrix elements $S(0,0,0,q),  S(2,2,0,q)$ and  $S(2,0,0,q)$ of
the static structure factor
$S(l_1,l_2,m,q)$ for $\phi$ = 0.62 
which is related to the total
correlation function $h(l_1,l_2,m,q)$ by 
\begin{eqnarray}
S(l_1,l_2,m,q) &=& \delta_{l_1,l_2} + \frac{\rho}{4 \pi} h(l_1,l_2,m,q)
\nonumber  \\ 
\underline{\underline{S}}(m,q) &=& \underline{\underline{1}}
+ \frac{\rho}{4 \pi} \underline{\underline{h}}(m,q) \nonumber  \\ 
\underline{\underline{S}}(m,q) &=& \left ( \underline{\underline{1}}
- \frac{\rho}{4 \pi} \underline{\underline{c}}(m,q) \right )^{-1} 
\end{eqnarray}
The q-axis has been stretched by a factor of $\sqrt[3]{X_0}$. It can be
clearly seen that a symmetry $X_0 \longrightarrow 1/X_0$ is 
not exactly valid at such high densities.

\subsection{Nematic instability}
Close to the nematic instability the matrix element
$S(2,2,m,q)$ of the static structure factor
develops a divergence at $q \longrightarrow
0$. This was already discussed in \cite{perera87} for results based on
the HNC (hyper-netted chain) closure relation.

\begin{figure}
\unitlength1cm
\epsfxsize=8cm
\begin{picture}(7,18)
\put(-1.2,12){\rotate[r]{\epsffile{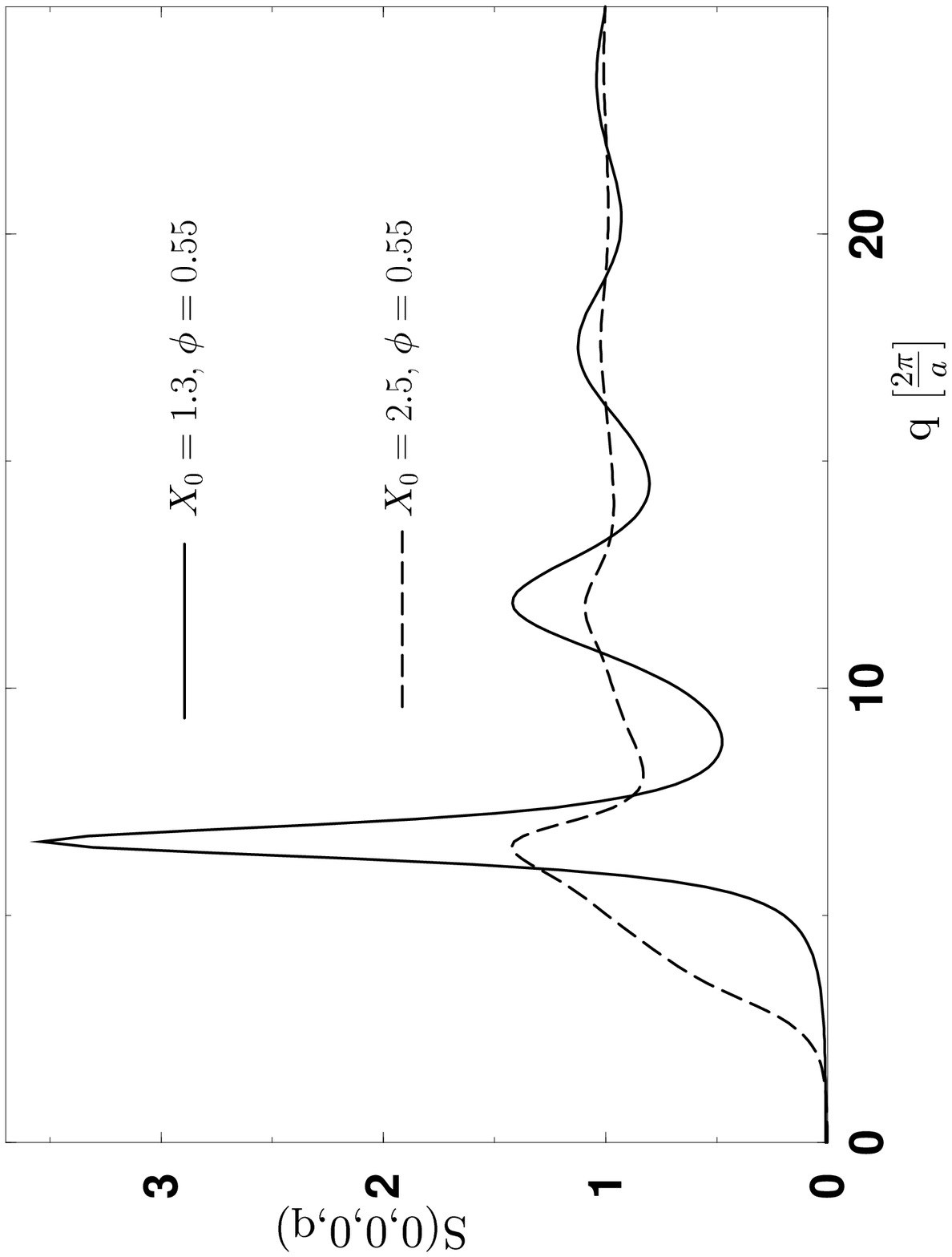}}}
\epsfxsize=8cm
\put(-1.2,6){\rotate[r]{\epsffile{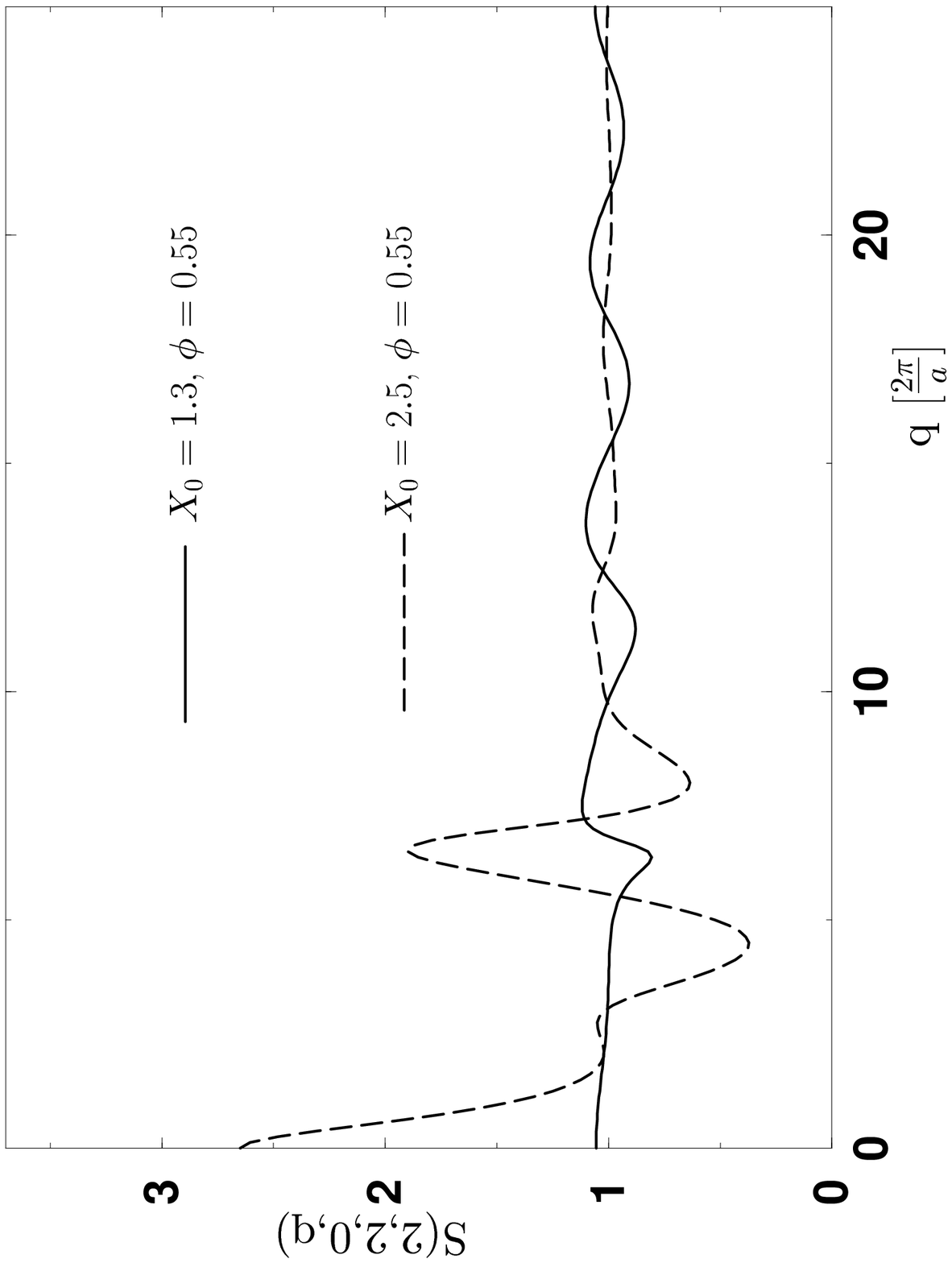}}}
\epsfxsize=8cm
\put(-1.2,0){\rotate[r]{\epsffile{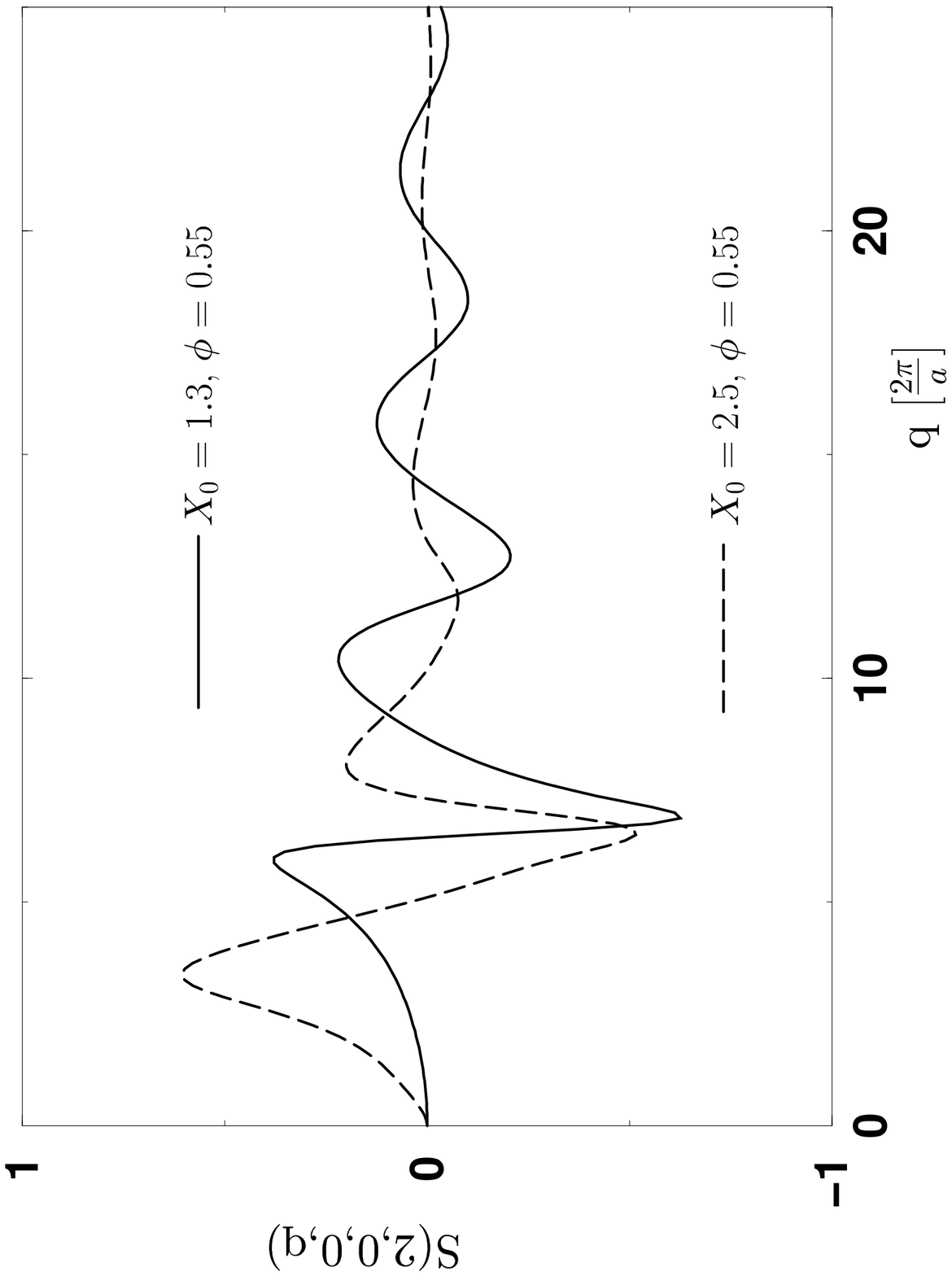}}}
\end{picture}
\caption{The static structure factors of two systems of ellipsoids
close to and far away from the 
nematic instability are
shown by comparing the
$S(l_1,l_2,m,q)$ components for $\phi=0.55$ for $X_0 = 1.3$ (far away
from the nematic instability) and for $X_0 = 2.5$ (close to the
nematic instability). In part (a) the
$S(0,0,0,q)$ is plotted in part (b) $S(2,2,0,q)$ and in part (c)
$S(2,0,0,q)$. Note that the $S(2,0,0,q)$ components vanish at $q=0$
due to the symmetries. 
}
\label{fig:nem}
\end{figure}

In fig. \ref{fig:nem}b such a precursor of a divergence is seen where
for $\phi = 0.55$ two systems $X_0=1.3$ (far from the nematic phase)
and $X_0 = 2.5$ (close to the nematic instability) are compared.
For  $X_0=1.3$ the $S(0,0,0,q)$ matrix element dominates as can be
seen from fig. \ref{fig:nem}a.
Note also from fig. \ref{fig:nem} that the $S(0,0,0,q)$ components and
the $S(2,2,0,q)$ seem to change their role when going from $X_0=1.3$
to $X_0=2.5$. The static structure for $X_0=1.3$ is dominated by the
$S(0,0,0,q)$ component while the orientational correlator $S(2,2,0,q)$
is small. This behaviour is reversed when looking at $X_0=2.5$. Here
the static structure is dominated by the orientational correlations
$S(2,2,0,q)$ while the center of mass $S(0,0,0,q)$ only shows a weak
structure.
Due to the fact that all the
non-diagonal elements of $S$ at $q \rightarrow 0$ vanish (as can be
seen in fig. \ref{fig:nem}c for $S(2,0,0,q)$) only
the $c(2,2,m,0)$ component governs the nematic instability and we get
as a condition for such an instability
\begin{equation}
\label{eq:nemcond}
\lim_{q \rightarrow 0} \left (
1 - \frac{\rho}{4 \pi} c(2,2,m,q) \right ) \longrightarrow 0 \;\;\;\; .
\end{equation}
This expression is the inverse of the Kerr
constant $K$ for non dipolar potentials.
\begin{equation}
K^{-1} = 
\lim_{q \rightarrow 0} \left (
1 - \frac{\rho}{4 \pi} c(2,2,m,q) \right ) \;\;\;\; .
\end{equation}

\begin{figure}
\unitlength1cm
\epsfxsize=10cm
\begin{picture}(7,16)
\put(-2.6,7){\rotate[r]{\epsffile{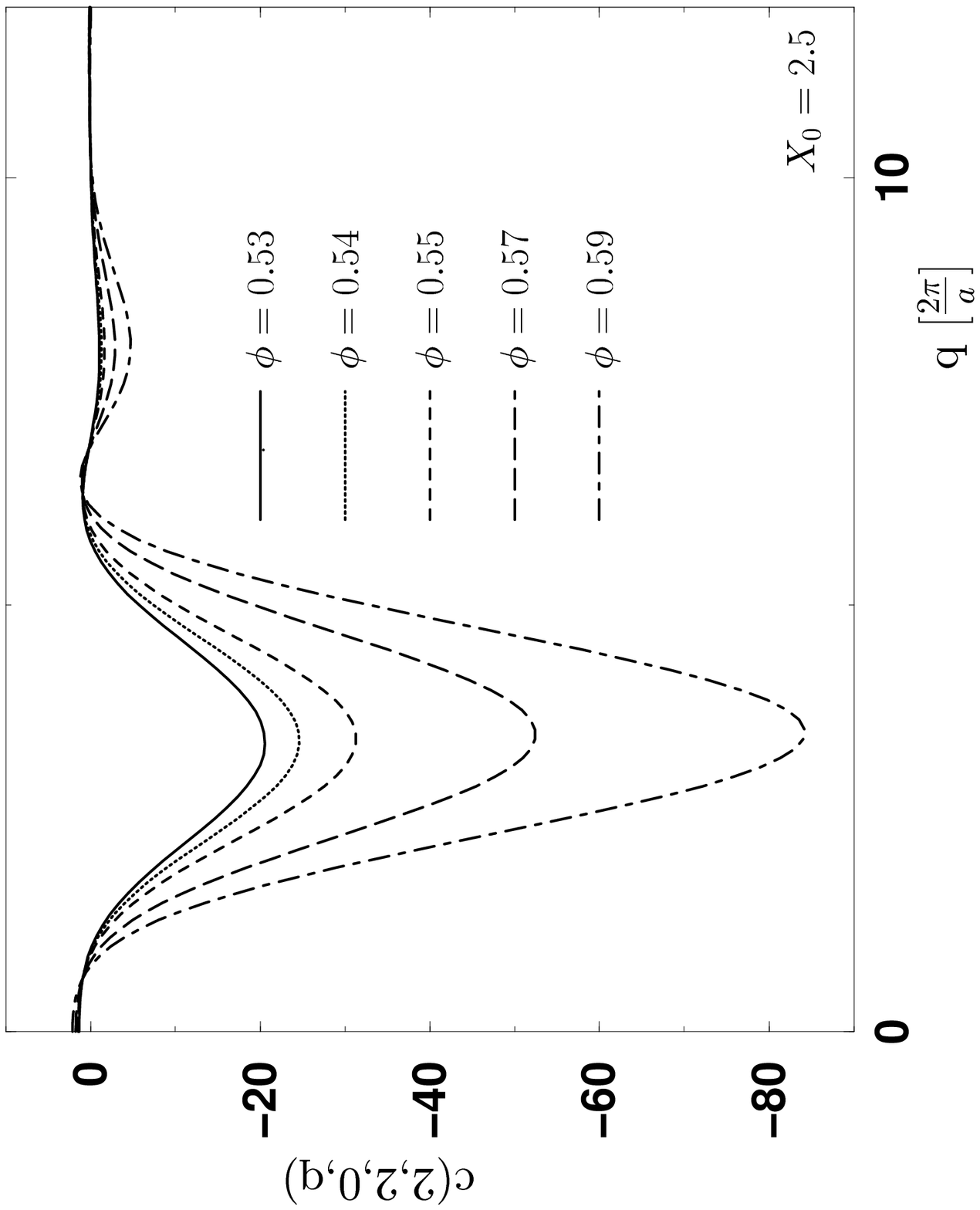}}}
\epsfxsize=10cm
\put(-2,0){\rotate[r]{\epsffile{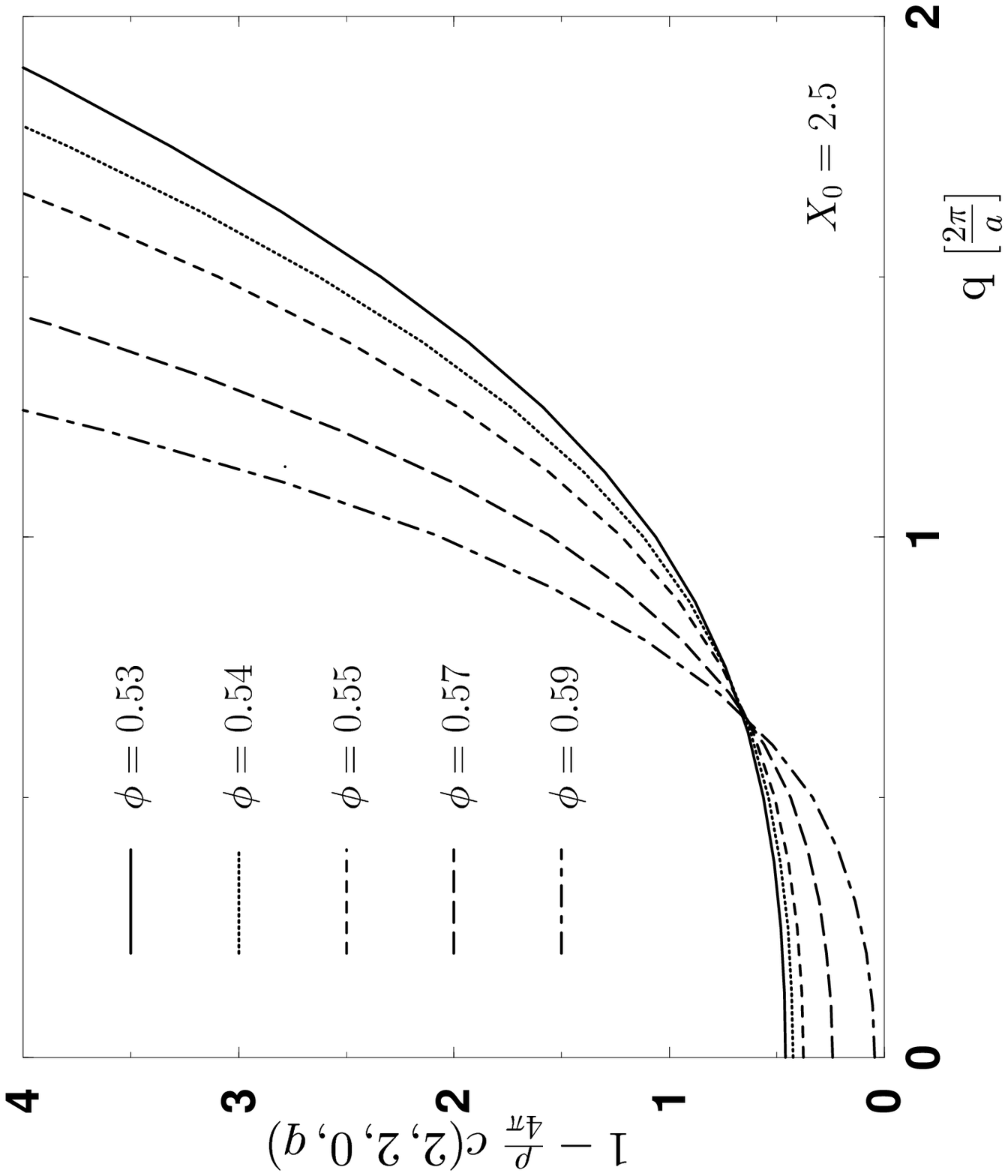}}}
\end{picture}
\caption{In (a) $c(2,2,m,q)$ for $q \longrightarrow 0$ 
and for different densities 
is plotted in (a). The curves for $\phi = 0.53, 0.54, 0.55$ were
obtained by a solution of the PY equations whereas the curves for
$\phi = 0.57, 0.59$ were obtained by applying a quadratic
extrapolation to higher densities to $c(2,2,0,q)$. The important part
for the nematic instability is plotted in (b) where the function
$1-\frac{\rho}{4 \pi} c(2,2,0,q)$ is drawn which becomes at $q=0$ the inverse
of $S(2,2,0,q=0)$.
}
\label{fig:nemc}
\end{figure}

A detailed analysis of the $q \rightarrow 0$ behavior therefore gives
us a condition for the nematic instability
as it is also discussed in a similar way in \cite{klapp96}
where dipolar fluids in the HNC approximation have been considered. The
instability is demonstrated in fig. \ref{fig:nemc}. In part (a) for
$X_0=2.5$ the function $c(2,2,0,q) $ is plotted for different
densities close to the nematic instability. The first three densities
$\phi = 0.53, 0.54, 0.55$ were the highest ones we could reach with the
numerical solution of the PY equation and the two higher densities
$\phi = 0.57, 0.59$ are quadratic extrapolations. In
fig. \ref{fig:nemc}b this is shown in greater detail where
$ 1-\frac{\rho}{4 \pi} c(2,2,0,q)$ is plotted which  
it becomes the
inverse of $S(2,2,0,0)$ at $q=0$. For X$_0$ = 2.5 The critical density
where $S(2,2,0,0)$ 
diverges is $\phi = 0.593$, according to such a quadratic extrapolation.

\subsection{Equilibrium phase diagram}

\begin{figure}
\unitlength1cm
\epsfxsize=10cm
\begin{picture}(7,8.5)
\put(-2.6,0){\rotate[r]{\epsffile{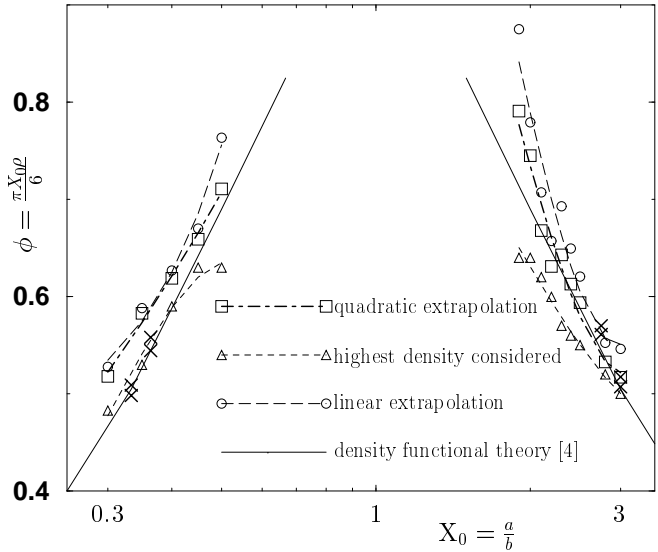}}}
\end{picture}
\caption{Isotropic to nematic instability as it arises from a Percus--Yevick
calculation. The highest densities considered are plotted with
triangles. From there an extrapolation was done to determine the
instability by $c(2,2,0,q \rightarrow 0) = 4 \pi / \rho $. This was
done with a linear (circles) and a quadratic (squares)
interpolation. For comparison we plotted the I-N-transition which
arises from density functional theory \protect\cite{groh97}. Also the
results from ref. \protect\cite{frenkel84} are plotted with black
crosses. 
}
\label{fig:phasediag}
\end{figure}

Using the above condition (eq. (\ref{eq:nemcond}))
we get the phase diagram for the
hard ellipsoids as it arises from the PY approximation as
shown in fig. \ref{fig:phasediag}. For all densities considered there
results a clear indication for a nematic instability. This phase boundary
still depends on the way the extrapolation to higher densities is done
(here linear or quadratic) but is in good agreement with other
works. For example density functional theory of Groh and Dietrich
\cite{groh97} are in a reasonable good agreement with our
results. However due to their 
approximation the $X_0$ to $1/X_0$ symmetry is exact
which is however clearly broken in the PY result. Also the results of Frenkel
et al. \cite{frenkel84} are along the same line. It also seems to be
that density--functional theory shows a better agreement with PY
theory for prolate ellipsoids that for oblate ones.



\section{Conclusion}

Orientational degrees of freedom in molecular systems can drive a
phase transition into an orientationally ordered nematic liquid
crystal phase. In principle integral equations have the ability to
describe a precursor phenomenon of such an orientational
transition. Until now this is well known for e.g. hyper-netted chain
(HNC) theory. In this work we demonstrate that also Percus--Yevick
theory gives a clear precursor of the nematic phase. We therefore were
able to calculate the equilibrium phase diagram of hard ellipsoids of
revolution.\\
The obtained phase diagram is in good agreement with density
functional theory \cite{groh97} and Monte Carlo simulations
\cite{frenkel84}.

\acknowledgements
M.L. thanks Mike Allen for making his results of computer simulations
available which made us quite confident about the quality of the
Percus--Yevick results.
We are further grateful to R.~Schilling for
valuable and helpful discussions and acknowledge
financial support from 
the Deutsche Forschungsgemeinschaft through
SFB 262. 


\end{document}